\documentclass{article}
\usepackage[final]{neurips_2024}
\usepackage[utf8]{inputenc} 
\usepackage[T1]{fontenc}    
\usepackage{hyperref}       
\usepackage{url}            
\usepackage{booktabs}       
\usepackage{amsfonts}       
\usepackage{nicefrac}       
\usepackage{microtype}      
\usepackage{xcolor}         
\usepackage{hyperref}
\usepackage{graphicx}

\title{Responsible Artificial Intelligence (RAI) in U.S. Federal Government : Principles, Policies, and Practices }

\author{%
   Atul Rawal, Katie Johnson, Curtis Mitchell, Michael Walton, and Diamond Nwankwo.   \\
  xD, U.S Census Bureau\\
  Washington D.C\\
  \texttt{atul.rawal@census.gov} \\
}

\begin{document}

\maketitle

\begin{abstract}

Artificial intelligence (AI) and machine learning (ML) have made tremendous advancements in the past decades. From simple recommendation systems to more complex tumor identification systems, AI/ML systems have been utilized in a plethora of applications. This rapid growth of AI/ML and its proliferation in numerous private and public sector applications, while successful, has also opened new challenges and obstacles for regulators. With almost little to no human involvement required for some of the new decision-making AI/ML systems, there is now a pressing need to ensure the responsible use of these systems. Particularly in federal government use-cases, the use of AI technologies must be carefully governed by appropriate transparency and accountability mechanisms. This has given rise to new interdisciplinary fields of AI research such as \textit{Responsible AI (RAI)}. In this position paper we provide a brief overview of development in RAI and discuss some of the motivating principles commonly explored in the field. An overview of the current regulatory landscape relating to AI is also discussed with analysis of different Executive Orders, policies and frameworks. We then present examples of how federal agencies are aiming for the responsible use of AI, specifically we present use-case examples of different projects and research from the Census Bureau on implementing the responsible use of AI. We also provide a brief overview for a Responsible AI Assessment Toolkit currently under-development aimed at helping federal agencies operationalize RAI principles. Finally, a robust discussion on how different policies/regulations map to RAI principles, along with challenges and opportunities for regulation/governance of responsible AI within the federal government is presented.   

\end{abstract}

\section{Introduction}

Artificial intelligence (AI) and machine learning (ML) have made tremendous progress since the early days of conceptual theories, to being integrated in almost every aspect of technological society. This progress has resulted in wide-spread application and adoption of data-driven artificial learning systems. With some systems capable of having very limited to almost zero human involvement/supervision, with the AI/ML systems making decisions based on the learned data. There is a crucial need to ensure the responsible use of these systems when they are utilized in sensitive applications such as healthcare and medicine \cite{arrieta2020explainable, goodman2017european}.  How can we ensure the trustworthiness of AI/ML systems, that they will be free of any inherent/implicit biases when making decisions? From Amazon's recruitment AI system which was biased against women, to Meta's advertisement AI system that was biased against race, gender and religion, there are plenty of real-world examples of AI failures that have resulted in harm to humans \cite{Dastin2018} \cite{Hao2019}. Another well known example of AI failures affecting human lives is the U.S health-care's bias against people of color \cite{Ledford2019}. 

These real world challenges and risks from AI/ML systems have resulted in calls for better trustworthy and accountable AI/ML systems. Industry, academia and government organizations need to ensure the utilization of AI/ML systems does not result in any potential harm from bias, safety, privacy and fairness. This has led to the creation of new laws and policies internationally and within the United States. Internationally, laws such as the European Unions, General Data Protection Regulation, which provides consumers with a “Right to Explanation” have aimed at protecting consumers from any harmful effects of irresponsible AI systems \cite{voigt2017eu}. Domestically, laws and policies have been created with the aim of making AI safe, trustworthy and responsible. Laws such as the U.S Algorithmic Accountability Act of 2019, dictates “assessments of high-risk systems that involve personal information or make automated decisions.” \cite{maccarthy2019examination} Executive orders (EOs) such as EO13960 and EO14410 are aimed at ensuring trust and safety of AI utilization within US federal agencies. 

Responsible AI (RAI) is the proposed solution to ensure that the AI systems used to make decisions were designed and built responsibly, and can be trusted by the end-users. RAI incorporates ethical aspects of AI/ML design and development. These principles when applied/practiced throughout the AI/ML system life-cycle can ensure that the systems are robust and trustworthy. Even though the field of AI/ML itself has been around for decades, there has only recently been an increase in the interest in responsible/ethical AI. (Fig \ref{fig:Scopus_RAI} (A)) highlights the increase in the number of yearly publications for \textit{responsible, ethical and trustworthy AI}. We see a rapid increase in the last decade, especially after 2015 when focus on making AI systems trustworthy and responsible garnered interest from governmental agencies such as DARPA with the creation of programs like the XAI program. 

Even though there are related surveys on RAI\cite{benjamins2019responsible, radanliev2024ethics, schiff2020principles, lu2022responsible, lu2024responsible, cheng2021socially}, which provide great overviews of RAI, a paper that provides a more comprehensive look at not just RAI development, but it's implementations and policies is needed,  especially from a regulatory perspective. Further, there is a lack of studies that highlight implementation of RAI principles at federal agencies. This position paper aims at providing a brief overview of RAI and regulatory policies associated with the responsible use of AI including EOs and federal memos. The main contributions of this position paper include a) an updated overview of RAI by focusing on the five principles, b) an overview of the current regulatory landscape related to RAI, c) use case examples of RAI practices at the Census Bureau(CB), d) RAI assessment toolkit currently under development at the CB, e) an open discussion of challenges that still remain within the field and perspectives on recommendations for addressing them.

\section{Overview of Responsible AI}

With the recent advancements in AI/ML, there have been associated challenges raising concerns about the implementation of AI/ML systems for sensitive applications such as healthcare and defense. These concerns have sparked new debate about the ethical principles required for the design, development, implementation and utilization. Here the concept of Responsible AI has provided principles/solutions to ensure the safe and trustworthy deployment of AI systems. Cheng et al., defined RAI as \textit{"intelligent algorithms that prioritize the needs of all stakeholders as the highest priority, especially the minoritized and disadvantaged users, in order to make trustworthy decisions. These obligations include protecting and informing users, preventing and mitigating negative impacts, and maximizing the long-term beneficial impact. (Socially) Responsible AI Algorithms constantly receive feedback from users to continually accomplish the expected social values"} \cite{cheng2021socially, barletta2023rapid}.

\begin{figure*}[ht]
    \centering
    \includegraphics[scale = 0.35]{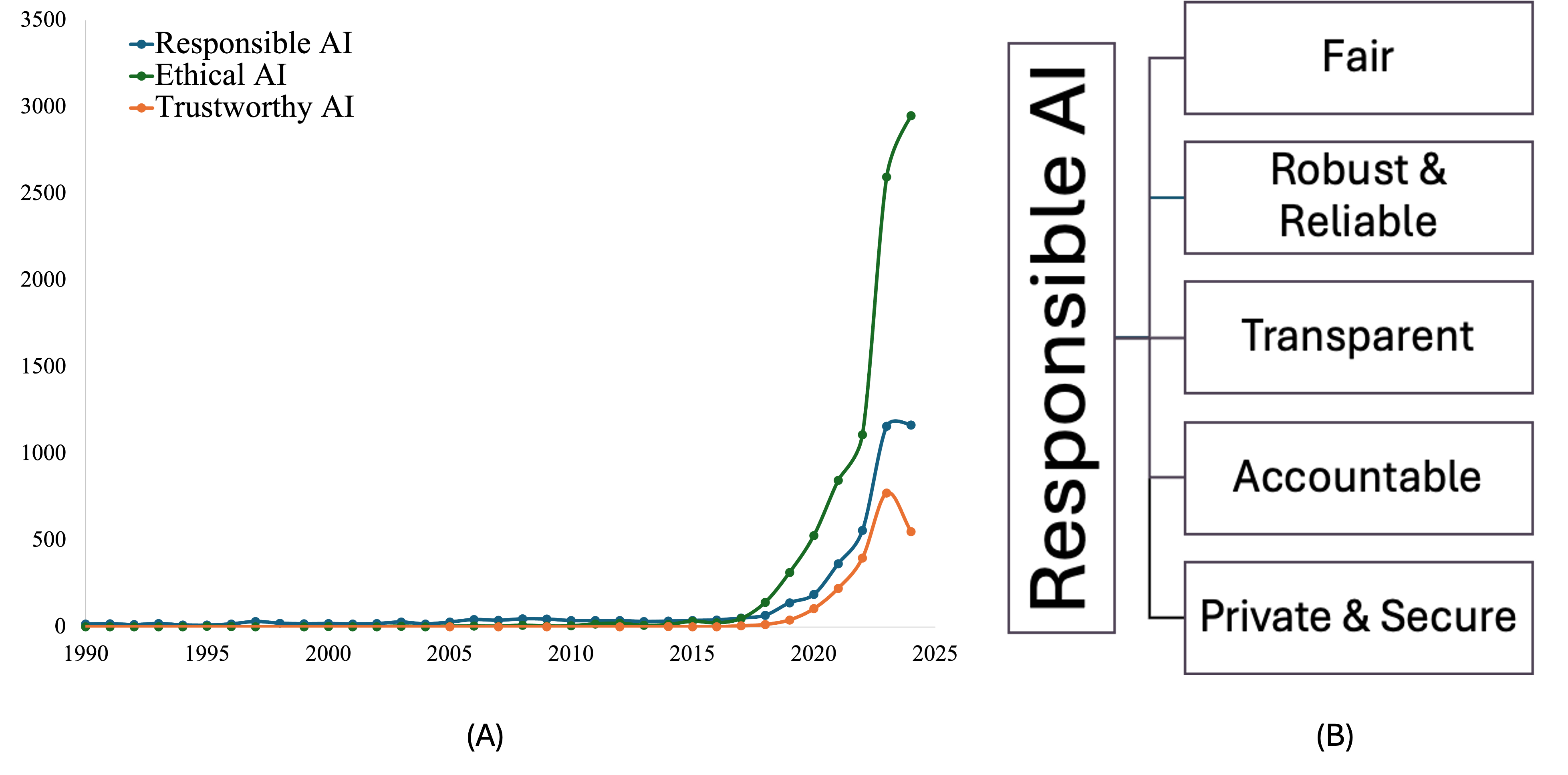}
    \caption{(A) Yearly publications for responsible, ethical and trustworthy AI. (Data derived from Scopus). (B) Five pillars of responsible AI}
    \label{fig:Scopus_RAI}
\end{figure*}

Numerous organizations in academia, government and industry have produced frameworks, guidelines, and policies related to ethical/responsible AI. These include Microsoft, Google, IBM, the Department of Defense, the National Institute for Standards and Testing (NIST),  University of Texas, and MIT \cite{lu2024responsible, cheng2021socially, barletta2023rapid}. A common theme across all these are the five concepts identified as  pillars/tenets of RAI as highlighted in Fig \ref{fig:Scopus_RAI} (B). These five concepts have been used by different organizations in a variety of forms to form the core concepts of RAI. While some organizations have termed different pillars together, others have made distinction between them or not included some within their pillars. In this section we provide a brief overview of the five pillars. 

\textit{Fair AI} has gained substantial interest in the past decade. However defining \textit{fairness} in terms of AI has been challenging as the concept of fairness can be subjective and mean different things depending on the context and societal priorities. In general fairness for AI aims to ensure the just, equitable, impartial, unbiased and fair design, development and deployment of AI/ML systems \cite{richardson2021framework, feuerriegel2020fair}.  Examples of harmful and unfair AI systems in the real world include META's AI system which was shown to have discriminatory results causing a legal settlement between the company and the Department of Justice that includes \textit{court oversight and regular review of its compliance}.

\textit{Reliability and robustness }of AI systems have been vital factors in their successful deployment. The aim here is to develop AI systems capable of achieving the highest levels of accuracy and reliability even under various degrees of uncertainty. Once deployed these systems must be evaluated on a regular basis to endure proper testing and assurance leading to effective operations across the entire life-cycle. \cite{dalrymple2024towards, ryan2020ai, kosheleva2024make} 

\textit{Transparency} for AI systems is closely related to Explainable AI (XAI). Rawal et al., defined explainability for for AI systems as \textit{"a set of processes or methods that ensures that the system to capable of allowing humans to comprehend its overall decision and reasoning. Explainability can be understood as a summary of the overall working features and calculations that produce the final system output."} \cite{rawal2021recent}. 

\textit{Accountable AI} aims at creating oversight for AI/ML systems to ensure control and accountability of the systems with humans. The NTIA's Artificial Intelligence Accountability policy states, \textit{"To be accountable, relevant actors must be able to assure others that the AI systems they are developing or deploying are worthy of trust, and face consequences when they are not."} In more layman terms accountability for AI refers to the expectation that design, development, and deployment of the AI/Ml system will be in compliance with all the policies/regulations and standards throughout it's life-cycle. It aims to create \textit{compliance, oversight, and enforcement} for AI/ML systems.

\textit{Privacy and security} are crucial part of trustworthy/ responsible AI, as even under uncertainty conditions such as perturbations and adversarial attacks, AI/ML systems must be robust and reliable.  Security refers to the protection of data integrity, confidentiality, and continuous functionality to the end users \cite{cheng2021socially}. Whereas privacy refers to the protection of the sensitive data used for model training or validation. Because sensitive data are protected by federal regulations such as Title and CIPSEA, data security for these types of data is an imperative condition of achieving data privacy. \cite{goellner2024responsible}

When combined together, these five concepts can ensure proper responsibility for AI/ML systems. However, similar to a table with a missing leg, RAI status cannot be achieved if any one of the five pillars are not implemented properly within the AI life-cycle. 

\section{RAI for federal agencies: policies and frameworks}

As mentioned multiple times throughout this article, AI/ML systems have seen tremendous progress within the past decade. However, laws and regulations to ensure the safe and responsible use of these systems have not kept pace, and are usually lagging behind. Especially in the U.S, AI regulations have not kept pace with regulations from places like the EU or Asia. During a 2023 US Senate hearing, the CEO of one of the most advanced AI companies in the nation/world, OpenAI, Mr. Sam Altman, while raising awareness of the exciting future/opportunities of AI also cautioned them about the dangers of AI when not done responsibly. He urged them to enact laws/policies for AI governance to ensure that these systems follow responsible principles/practices throughout their entire life-cycle. 

\begin{figure*}[ht]
    \centering
    \includegraphics[scale = 0.42]{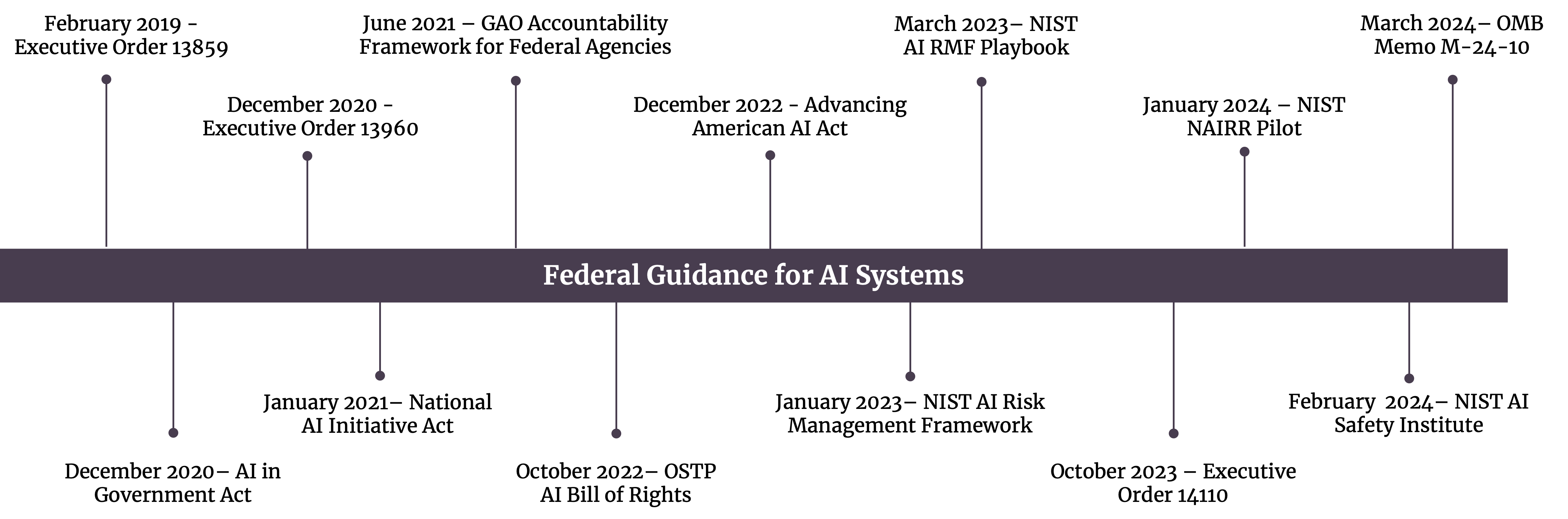}
    \caption{A timeline of AI related guidance/policies from the U.S government.}
    \label{fig:Timeline}
\end{figure*}

AI governance is aimed at ensuring risk minimization for potential harms from privacy leak, bias, discrimination, and unfair AI decisions. It can help build trust with consumers that their use of AI/ML systems will not result in harm to them. Towards this cause various federal, commercial and international organizations have put forth governance principles and guidelines aimed at supporting AI/ML systems life-cycle to mitigate and reduce the risks of not adopting AI systems responsibly. This section provides an overview of the different frameworks and regulations within the US aimed at ensuring the responsible utilization of AI [Fig \ref{fig:Timeline}]. 

\subsection{Executive Actions}

Three Executive Orders (Executive Orders 13859, 13960, \& 14110) have been issued outlining guiding principles and priorities for responsible development and usage of AI technologies. In support of these EOs, the Office of Management and Budget (OMB) memo on “Advancing Governance, Innovation, and Risk Management for Agency Use of Artificial Intelligence” directs federal agencies to establish governance protocols and adopt appropriate risk management practices for AI innovation. 

Executive Order 13859 "Maintaining American Leadership in Artificial Intelligence" issued on February 11th 2019 outlined the aims for strengthening American leadership in AI \cite{EO13859}. It outlined five key efforts to highlight American influence in AI: a) increasing investments in AI research, b) prioritizing federal investments in AI using federal AI computing and data resources, c) setting technical AI standards, d) increasing the AI workforce, and e) engaging with international allies on AI collaborations. The EO was one of the first ever to substantially increase funding for AI related investments, especially for non-defense AI R\&D, including more than \$800 million for the National Science Foundation, \$125 million for DoE, \$50 million for NIH, and \$100 million for Agriculture and Food Research Initiatives. It also invested \$1 billion in establishing 12 AI and Quantum Information Science (QIS) R\&D centers such as the NSF National AI Research Institutes. 

Executive Order 13960 “Promoting the Use of Trustworthy Artificial Intelligence in the Federal Government”, issued on December 3, 2020, outlines two major requirements for federal agencies \cite{EO13960}. The first requirement is that federal agencies adhere to nine principles when “designing, developing, acquiring, and using AI in the Federal Government.”  The nine principles are: (a) lawful and respectful of Nation’s values, (b) purposeful and performance-drive, (c) accurate, reliable, and effective, (d) safe, secure, and resilient, (e) understandable, (f) responsible and traceable, (g) regularly monitored, (h) transparent, and (i) accountable. The second requirement established annual AI use-case inventories for each federal agency for non-classified and non-sensitive uses cases of AI. Also, the Executive Order applied to both existing and new uses of AI. While the intention of the EO had been well received by federal agencies, it's implementation has not been completely achieved. A 2023 GAO report stated that \textit{"federal agencies have taken initial steps to comply with AI requirements in executive orders and federal law; however, more work remains to fully implement these."} For the requirement of publishing agencies' AI use-cases, the study identified \textit{inaccurate and incomplete} data from different inventories. Based on these results the GAO made 35 recommendations to 19 federal agencies, including OMB, to fully implement the federal requirements related to AI. These recommendations included a) requiring 15 agencies to update the AI use case inventories to include the required information and its alignment with the guidance, b) requiring OMB, OSTP and OPM to implement AI requirements with implications spanning the entire federal government, and c) implementation of  AI requirements at 12 agencies with federal policy/guidance. 

Executive Order 14110 “Safe, Secure, and Trustworthy Development and Use of Artificial Intelligence”, issued on October 30, 2023, outlines eight guiding principles and priorities to advance and govern the use of AI: (a) ensure safe and secure AI technology; (b) promote responsible innovation, competition, and collaboration; (c) support American workers; (d) advance equity and civil rights; (e) protect American’s interests and consumer protection; (f) protect privacy and civil liberties; (g) manage risks of the federal government’s use of AI; and (h) strengthen U.S. leadership abroad, safeguarding ways to develop and deploy AI technology responsibly \cite{EO14110}. Executive Order 14110 also includes over 100 actions that span more than 50 federal entities. The EO puts a large portion of the responsibilities for the actions to be taken on the DoC/NIST, requiring NIST to collaborate with the Secretaries of Energy, Homeland Security and other relevant agency heads to establish industry-leading best practices for trustworthy AI systems. It also requires NIST to establish guidelines for red-teaming tests to ensure the robustness, safety, security and trustworthiness of these systems. Other actions include requiring companies involved in the development of dual-use foundational models for federal agencies to provide data on training, development and deployment of foundation models such as model weights and performance results from red-team testing. To tackle the misuse of generative AI features, the EO dictates the \textit{"identification of existing standards, tools, methods, and practices for disclosing generated content. Including methods for authenticating content and tracking provenance, labeling synthetic content, detecting synthetic content, and preventing generative AI from producing child sexual abuse material or non-consensual intimate imagery of real people.}" It also directed NIST to launch the National AI Safety Institute and  the National AI Research Resource (NAIRR) pilot program which has been implemented and shown early success. While it is still fairly early to make judgments on the success of the EO or its compliance/implementation by federal agencies, actions taken so far by NIST and other federal agencies do signal a significant determination to ensure the responsible use of AI within the federal government. 

The OMB M-24-10 memorandum provides directives for federal agencies on improving governance, innovation, and risk management in their use of Artificial Intelligence (AI) \cite{OMB2410}. It is aimed at ensuring that AI is used in a safe/secure manner, and it aligns with ethical standards, particularly concerning public safety and civil rights. The major directives are aimed towards: a) AI governance, b) RAI innovation, c) AI risk management. For AI governance the memo directs agencies to designate a Chief AI Officer (CAIO) along with an AI Governance body. In one of the most direct and impactful regulations to date, the memo also directs agencies to "\textit{submit to OMB and post publicly on the agency’s website either a plan to achieve consistency with this memorandum, or a written determination that the agency does not use and does not anticipate using covered AI. Agencies must also include plans to update any existing internal AI principles and guidelines to ensure consistency with this memorandum,  OMB will provide templates for these compliance plans.}" For RAI innovation the memorandum dictates agencies to increase RAI adoption capacity and transparency by sharing of AI models, code and data. Additionally, each CAIO must also develop an enterprise RAI strategy which should include increased access to AI resources such as data, infrastructure, and cybersecurity. Finally, for AI risk management the memorandum establishes new requirements and recommendations to address AI risks that can impact public rights and safety.  Agencies must follow minimum practices for managing AI risk that impacts public safety and civil rights. This includes conducting risk assessments, ongoing monitoring, and providing public notice of AI usage.  Agencies must evaluate the potential benefits and risks of AI, ensuring that benefits outweigh risks before deployment. The memo applies to all federal agencies as defined under the AI in Government Act of 2020, with specific exclusions for certain national security systems and regulatory actions. Agencies are required to meet various deadlines, including designating a CAIO within 60 days of the memo’s issuance and achieving compliance with the minimum practices for safety-impacting and rights-impacting AI by December 1, 2024. The memo establishes one of the first mandatory  frameworks for federal agencies to responsibly innovate with AI while managing risks, setting a precedent for future AI governance both within the government and potentially influencing private sector practices.

The AI Bill of rights, released in October 2022 identified five principles for responsible deployment of AI systems in the U.S \cite{AIbill}. While voluntary, the guidelines are intended to ensure the safety and privacy of consumers who utilize AI systems. The guidelines highlight the need for safe and effective AI systems by ensuring that designers, developers and deployers of AI systems are committed to protecting the consumers against any harmful or discriminating effects of AI predictions/decisions. The blueprint lists the following as \textit{"five principles and associated practices to help guide the design, use, and deployment of automated systems to protect the rights of the American public in the age of artificial intelligence}": a) safe and effective systems, b) algorithmic discrimination protections, c) data privacy, d) notice and explanations, and e) human alternatives, consideration and fallback. These guidelines are also intended specifically for automated systems with potential to impact the rights, opportunities of the American public and their access to resources/services. 

\subsection{AI Risk Management Frameworks}

The National Institute of Standards and Technology (NIST) AI Risk Management Framework (RMF) and associated RMF Playbook provide resources for organizations developing, deploying or using AI systems to manage risks and promote trustworthy and responsible usage \cite{925786}. It is designed and developed as  unbiased/neutral guidelines to aid organizations evaluate and manage AI related  risks. The framework was created as a response to the ever-evolving landscape of risks and challenges associated with the AI adoption across industries and federal agencies. It provides a comprehensive and structured view of evaluating and mitigating AI related risks. The framework defines seven core principles in evaluating the trust and reliability of an AI system; a) safety b) validity and reliability, c) security and resiliency, d) improved/enhanced privacy, e) transparency and accountability, f) interpretability and explainability, and g)fairness and bias mitigation. The framework does highlight that simply addressing each of the principles individually will not result in a trustworthy AI system. Rather, \textit{"creating trustworthy AI requires balancing each of these characteristics based on the AI system’s context of use....Neglecting these characteristics can increase the probability and magnitude of negative consequences. Trade-offs are usually involved, rarely do all characteristics apply in every setting, and some will be more or less important in any given situation."} The framework provides four core functions to ensure trustworthy AI; a) govern, b) map, c) measure, and d) manage. 

Similarly, The U.S. Government Accountability Office (GAO) “Accountability Framework for Federal Agencies and Other Entities” provides key accountability practices in governance, data performance and monitoring to help federal agencies use AI responsibly \cite{GAO}. The document highlighted that while federal policies and guidelines have focused on ensuring AI is responsible, equitable, traceable, reliable, and governable, evaluations and assessments of AI systems using third-party audits are necessary. Additionally, it also highlighted a shortage of AI-related expertise/talent in the federal workforce needed to accelerate the adoption of AI. These documents provide useful clarification and detailed procedures for operationalizing the principles laid out in recent Executive Orders, OMB Memo and other AI policy documents. 

\section{RAI Practices - Census Bureau use-case examples}

The Census Bureau (CB) collects data on almost every aspect of the nations commerce activities such as trade, economy, demography, and geography. This data is provided by the department to provide a snapshot of the country’s population size and growth along with the characteristics of the economy. It also performs vital statistical analysis that informs an accurate portrait of the population. All this information is collected through numerous surveys such as the American Community Survey (ACS), the economic census, and the decennial census. The decennial census, as the name suggests is performed every ten years ending in zero to count each resident of the country and where they live on April 1. This is mandated by the US Constitution to determine the apportion of the House of Representatives among the states. It also provides valuable data for a number of other applications such as informing policies for the distribution of hundreds of billions of dollars in federal funds for state and local governments. With the boom in utilization of AI/ML systems across federal agencies, CB has also adopted AI for various applications ranging from research to operations. To ensure the use of AI aligns with the policies and regulations set forth by the government, CB is prioritizing RAI practices within it's use of AI. Here we share examples of a few projects where RAI principles are being practiced and other projects aimed at helping CB utilize RAI principles. 

\subsection{Model card generator \& AI registry}

\textit{Model cards} are a form of documentation that describe a model's data, architecture, performance, and intended users and use cases. They are intended to be comprehensible by a wide audience, and serve to reduce bias and increase transparency by documenting various qualities about a model and its origins. The CB's model card generator feature consists of a web form and associated JavaScript to create a markdown file populated with information a user provides in the web form. The form's fields include sections requiring descriptions of the model's use cases, its architecture, what data the model was trained on and how training performance was evaluated, efforts to detect and mitigate bias in the data and model performance, and information on the model's compliance with existing laws and regulations. The form design and accessibility conform to requirements from the United States Web Design System (USWDS). When a user has completed filling out the proper information, the application creates a markdown file that their browser prompts the user to download. Importantly, this is all done within the user's browser, meaning that no information contained in the model card is sent to a remote server. 

The \textit{AI Registry} aims to build trust in AI/ML by building a centralized register of AI/ML models used within the CB to highlight comprehensive oversight and governance.  Due to the ever-increasing need for AI/ML governance, the registry allows for a centralized repository of these models to evaluate them for accountability, which includes explainability, fairness, bias identification \& mitigation and compliance with the various policies and regulations. It allows for an effective and more efficient management with a centralized library for all AI projects with aims to store information on all the AI/ML models in one repository providing a top-down approach to transparency of the model’s attributes including bias, fairness and risk management. The repository enables AI teams to track, manage and share the unique attributes of each model across different teams and stakeholders looking to utilize similar (if not the same) models for their own applications. By creating and maintaining such a repository CB is at the forefront of accountable AI/ML governance. Teams/divisions using AI/ML products within their research or duties are be able to submit their models directly to the register to highlight accountability and compliance. This aligns the CB's adoption of AI/ML initiatives with responsible AI practices across the bureau by populating the repository with various AI/ML models. For each model deposited into the register accompanying information on the model includes information related to its data, data source, application, fairness, bias identification/mitigation methods (if applicable), explainability capabilities, compliance with regulatory policies and basic technical details about the model algorithm.  The main value add of the registry along with transparency and reusability are effective AI/ML management and governance. 

\subsection{RAI assessment toolkit}

Statistical agencies have been exploring the use of AI to produce statistics for many years. As the sophistication and availability of AI tools increases, the CB anticipates an increase in the use of AI to support statistics production, from mission enabling activities through the entire lifecycle of data collection and dissemination. The use of AI introduces exciting opportunities and legitimate concerns about the autonomous nature of these tools and their known shortcomings, including algorithmic bias and reputational harm due to mistrust in AI, that may impact data quality and survey participation negatively. To support the highest standards of statistical quality when using AI, the team at the CB has partnered with the General Services Administration (GSA) to develop a Responsible AI (RAI) assessment toolkit for AI’s use, with a focus on supporting statistics production. Modeled after the DoD’s RAI toolkit, this toolkit will focus on RAI principles and risk management for statistical agencies that work with sensitive data such as Title or CIPSEA data.

It is aimed at helping federal teams involved in the design/development/deployment of AI/ML systems ensure compliance with different RAI principles and federal guidelines/policies. It provides a seamless process for translating high-level principles, policy, and guidance into concrete metrics and technical processes. It is designed to be a centralized repository for assessment of RAI principles and polices, along with an available open-source tools designed for different RAI principles. It provides a bottom-up approach to transparency and trust by allowing AI/ML teams to identify, track, manage and improve the alignment of their AI/ML projects to RAI principles and best practices. It is a web-based UI application aimed at providing a suite of tools for data and model evaluation. The toolkit is modeled after the DoD's RAI Toolkit, however it is differentiated by the intended users and focus on regulations, as the toolkit is designed for federal AI teams that work with protected data such as Title and CIPSEA. The assessment tool allows designers/developers to investigate the alignment of their AI systems to the RAI principles outlined in this paper and the different federal guidelines and requirements outlined in the various EOs and memos. A RAI assessment will direct users to the available tools that are applicable to the AI model/data to conduct the RAI alignment evaluation. For example, AI teams working with Title 13 data are required by law to protect the privacy of data and model. Towards this goal, AI teams can utilize the RAI assessment toolkit to investigate applicable regulations. The assessment toolkit would then suggest and highlight the requirements for compliance with applicable privacy regulations such as the OMB 24-10 memo, EO 14110, and RAI principle of \textit{"privacy and security"}. The toolkit would then provide a list of available tools/resources such as SMPCs, homomorphic encryption and other PETs technologies to protect the privacy of the data and model. As an added function, the toolkit also highlights applicable risk-management resources/tools that are available for ensuring model and data specific risk identification and mitigation strategies. As a second example, if an income prediction project team were aiming to ensure alignment with the transparency/explainabilty portion of the RAI principle as required by EO 13960 and 14110, the RAI assessment toolkit will provide solutions towards this by providing a list of  available tools/resources for post-hoc explanations such as SHAP and LIME. The list of tools would help with the teams with generating both local and global explanations for the predictions made by the model, which would ensure alignment with the transparency principle and policy. 

\begin{figure*}[ht]
    \centering
    \includegraphics[scale = 0.55]{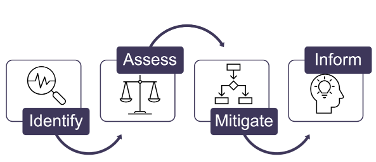}
    \caption{An overview of the RAI assessment toolkit framework.}
    \label{fig:toolkit}
\end{figure*}

\section{Discussion \& Conclusion}

As the reader might have noticed, AI systems have made quite the buzz in the past decade. For the federal government especially, AI systems can bring remarkable capabilities for a plethora of applications from automation to research. However, as the quote embedded in pop culture and Marvel history states, \textit{"with great power, comes great responsibility"}. For federal agencies to utilize AI solutions to the full potential, RAI practices must be embedded as core characteristics of AI systems, rather than post-hoc capabilities. Especially when these systems utilize sensitive information such as title protected data. To ensure the responsible and trustworthy use of AI, the federal government has provided guidance and policies to help federal teams align the design, development and deployment of AI systems with responsible/trustworthy practices. 

All the policies and regulations for AI have aimed to address the five pillars of RAI. EOs 13960 and 14110 address them through the following requirements. For fairness, EO 13960 specified the need for AI design, development and deployment to exhibit "\textit{due respect for our Nation's values...including those addressing privacy, civil rights, and civil liberties}." Whereas EO 14110 requires the fair deployment of AI stating that \textit{"AI should not be deployed in ways that undermine rights."} Towards reliability and robustness both EO 13960 and 14110 mandate that "\textit{AI is consistent with the use cases for which that AI was trained, and such use is accurate, reliable, and effective}" (EO 13960) and "\textit{requires robust, reliable, repeatable, and standardized evaluations of AI systems,}" (EO 14110). Transparency on the federal use of AI is also mandated by both the EOs, where EO 13960 required federal agencies to make publicly available the AI use-case inventory for non-classified and non-sensitive applications of AI, and EO 14110 provided further guidance for OMB to update instructions for reporting on agencies' use-cases and their risk management. Both EOs describe accountability for AI systems. In EO 13960, agencies are accountable "\textit{for implementing and enforcing appropriate safeguards for the proper use and functioning of their applications of AI, and shall monitor, audit, and document compliance with those safeguards.}" EO 14110 states that "\textit{It is necessary to hold those developing and deploying AI accountable to standards that protect against unlawful discrimination and abuse, including in the justice system and the Federal Government.}" Both EOs specify action towards privacy and security. EO 13960 states that "\textit{Agencies shall ensure the safety, security, and resiliency of their AI applications, including resilience when confronted with systematic vulnerabilities, adversarial manipulation, and other malicious exploitation.}" EO 14110 requires updated software-development lifecycle standards to protect "\textit{personally identifiable information, including measures to address AI-enhanced cybersecurity threats in the health and human services sector.}" Furthermore, EO 14110 includes the development of "\textit{a Secure Software Development Framework to incorporate secure development practices for generative AI and for dual-use foundation models.}"  

Federal regulations and guidelines in forms of EOs, memos and voluntary frameworks have outlined the required characteristics/properties that AI systems need to achieve to attain trustworthiness and responsible utilization. These guidelines have mandated requirements such as the publication of AI-inventory for federal agencies, created new national institutes to provide technical support and guidance for utilizing AI systems in trustworthy and responsible manner, and highlighted the need for an AI workforce for the federal government to keep up with industry partners such as Google \& Microsoft. These policies have brought about some of the first regulations aimed towards the responsible use of AI. However, there are still glaring challenges that cause concern including; a) uncertainties for setting explicit definitions of AI, b) risk determination and assessments, c) assigning responsibilities for different aspects of AI, d) fostering an innovation ecosystem without being perceived as "over-regulating", e) methods/techniques for evaluating the harms/failures of the defined AI systems, f) enhancing the federal AI capabilities with an enhanced and refreshed AI skilled/ready workforce, and g) keeping pace with an ever-evolving AI landscape where newer and more advanced models are released faster than regulations can keep-up with.

All of these challenges are further amplified by the recent rise of foundational models including LLMs that provide advanced new capabilities for the federal government. The black-box, non-open source and general purpose nature of foundation models can both amplify existing risks and pose new ones for regulators. For example, as highlighted in EO 14110, the identification and mitigation of synthetic materials such as images, videos, audio for watermarking purpose creates a monumental challenge when generative models are capable of creating almost real-like materials. Licensing is another challenge enhanced by the recent rise of LLMs, which can stifle innovation by limiting access to these models. The multi-modal, multi-application capabilities of LLMs also pose a regulatory challenge as the same models can be used for multiple applications ranging from rights-impacting tasks such as hiring, to mundane tasks such as automating filing. This also creates challenges for assessing the potential risks for these models when the applications of these models by the end-users vary so much. 

To address some of these challenges, a bottom-up/grass-roots approach must be applied at federal agencies. As the domain for different agencies varies, the AI related challenges and risks also vary amongst different agencies. For example, the Food \& Drug Administration does not have the same challenges related to privacy and security of Title 13 data as the Census Bureau does. Therefore, solutions for agency/domain specific AI risks must be tackled at the federal "grass-roots" level at each agency where the subject matter expertise lies. This is highlighted in this position paper with the AI projects/solutions being utilized at the CB which align with the RAI principles and the different EOs and frameworks. The model card generator is an excellent source of promoting documentation, which can be utilized for the AI registry which promotes transparency for the bureau's use of AI/ML. The RAI assessment toolkit provides a solution for the evaluation and assessment of domain specific AI systems utilizing title protected datasets. 

For the US federal government to catch up with other international partners such as the European Union, the United Kingdom, India, and China, AI-related regulations must be mandated not just at the federal level but also the industry/consumer level. Since most of the policies/guidance have been in forms of EOs, these face the potential of being revoked/removed when the administration changes. Therefore AI policies and regulations must be codified into law by Congress within the US to ensure their sustainable and responsible use.  This position paper provided an overview of RAI principles along with US regulations/policies governing the responsible/trustworthy use of AI, and highlighted projects at the CB that emphasize the commitment to RAI principles.

\bibliographystyle{IEEEtran}
\bibliography{references.bib}

\end{document}